\documentclass[12pt]{article}
\usepackage{graphicx}
\usepackage{hyperref}
\usepackage{amsmath}
\usepackage{amscd,amssymb}

\newcommand{\cF}{\mathcal{F}}

\newcommand{\cG}{\mathcal{G}}

\def\cR{{\mathcal R}}

\def\cB{{\mathcal B}}

\def\cK{{\mathcal K}}
\def\cD{{\mathcal D}}

\def\cP{{\mathcal P}}

\def\cL{{\mathcal L}}

\def\cF{{\mathcal F}}

\def\cY{{\mathcal Y}}
\def\cG{{\mathcal G}}

\def\mg{{\mathfrak g}}

\def\ml{{\mathfrak l}}
\def\ms{{\mathfrak s}}
\def\mg{{\mathfrak g}}

\newtheorem{remark}{Remark}[section]

\def\e{{\,\rm e}\,}

\newcommand{\beq}{\begin{eqnarray}}
\newcommand{\eeq}{\end{eqnarray}}
\numberwithin{equation}{section}

\begin{document}


\begin{center}
{\large\bf Multipartite Generating Functions and
Infinite Products for Quantum Invariants}
\end{center}

\vspace{0.1in}

\vspace{0.1in}

\begin{center}
{\large
A. A. Bytsenko $^{(a)}$
\footnote{E-mail: aabyts@gmail.com} and
M. Chaichian $^{(b)}$
\footnote{E-mail: masud.chaichian@helsinki.fi}}

\vspace{5mm}

$^{(a)}$
{\it
Departamento de F\'{\i}sica, Universidade Estadual de
Londrina\\ Caixa Postal 6001,
Londrina-Paran\'a, Brazil}

\vspace{0.2cm}
$^{(b)}$
{\it
Department of Physics, University of Helsinki\\
P.O. Box 64, FI-00014 Helsinki, Finland}

\end{center}

\vspace{0.1in}

\begin{abstract}
We show that multipartite generation functions can be written in terms of the Bell polynomials (known as Fa\`a di Bruno's
formula) and the Ruelle spectral functions, whose spectrum is encoded in the Patterson-Selberg function of the hyperbolic 
three-geometry. We derive an infinite-product formula for the Chern-Simons partition functions and analyze appropriate
q-series which leads to the construction of knot invariants. With the help of the Ruelle spectral functions symmetric and 
modular properties in infinite-product structure can be described.
\end{abstract}

\vspace{0.1in}

\begin{flushleft}
\\
\vspace{0.3in}
\end{flushleft}

\begin{flushright}
\emph{Dedicated to the memory of our friend and colleague, Petya Kulish}
\end{flushright}

\newpage

\tableofcontents

\section{Multipartite generating functions}

Let us consider, for any ordered $m$-tuple of nonnegative integers not all zeros,
$(k_1, k_2, \ldots ,k_m)={\overrightarrow{k}}$
(referred to as "$m$-partite" or {\it multipartite} numbers), the
(multi)partitions, i.e. distinct representations of $(k_1, k_2, \ldots ,k_m)$ as sums
of multipartite numbers. Let us call
${\mathcal C}_-^{(z; m)}({\overrightarrow{k}}) = {\mathcal C}_-^{m}(z;k_1, k_2 , \cdots , k_m)$
the number of such multipartitions, and introduce in addition the symbol
${\mathcal C}_+^{(z; m)} ({\overrightarrow{k}})=
{\mathcal C}_+^{(m)}(z;k_1, k_2 , \cdots , k_m)$.
Their generating functions are defined by \cite{Andrews1}
\begin{eqnarray}
{\mathcal F}({z;X}) &: = & \prod_{{{\scriptstyle k_1\geq 0, \ldots, k_m\geq 0,}
\atop\scriptstyle k_1+\ldots+k_m> 0}}\left( 1- z x_1^{k_1}x_2^{k_2}
\cdots x_m^{k_m}\right)^{-1}
= \sum_{{\overrightarrow{k}}\geq 0}{\mathcal C}_-^{(z;m)}({\overrightarrow{k}})
x_1^{k_1}x_2^{k_2}\cdots x_m^{k_m}\,,
\label{PF1}
\\
{\mathcal G }(z;X) &: = & \prod_{{{\scriptstyle k_1\geq 0, \ldots, k_m\geq 0,}
\atop\scriptstyle k_1+\ldots+k_m> 0}}\left( 1 + z x_1^{k_1}x_2^{k_2}
\cdots x_m^{k_m}\right)
= \sum_{{\overrightarrow{k}}\geq 0}{\mathcal C}_+^{(z; m)}({\overrightarrow{k}})
x_1^{k_1}x_2^{k_2}\cdots x_m^{k_m}\,.
\label{PF2}
\end{eqnarray}
Therefore,
\begin{eqnarray}
{\rm log}\, {\mathcal F}(z;X) & = & - \sum_{{{\scriptstyle k_1\geq 0, \ldots, k_m\geq 0,}
\atop\scriptstyle k_1+\ldots+k_m> 0}}{\rm log}
\left(1- z x_1^{k_1}x_2^{k_2}\cdots x_m^{k_m}\right)
=  \sum_{{\overrightarrow{k}}\geq 0} \sum_{n=1}^\infty \frac{z^n}{n}
x_1^{nk_1}x_2^{nk_2}\cdots x_m^{nk_m}
\nonumber \\
& = &
\sum_{n =1}^\infty \frac{z^n}{n}
(1-x_1^n)^{-1}(1-x_2^n)^{-1} \cdots (1-x_m^n)^{-1}
\nonumber \\
& = &
\sum_{n=1}^\infty \frac{z^n}{n}
\prod_{j= 1}^m (1-x_j^n)^{-1}.
\end{eqnarray}
Finally, ${\rm log}\,{\mathcal G}(-z;X)  =  {\rm log}\,{\mathcal F}(z;X)$.
Let $\beta_m(n):= \prod_{j=1}^m(1-x_j^n)^{-1}$, then
\begin{eqnarray}
\!\!\!\!\!\!\!\!\!\!\!\!\!
{\mathcal F}(z;X) & = &
\sum_{{{\scriptstyle k_1\geq 0, \ldots, k_m\geq 0,}
\atop\scriptstyle k_1+\ldots+k_m> 0}}{\mathcal C}_-^{(z; m)}({\overrightarrow{k}})
x_1^{k_1}x_2^{k_2}\cdots x_m^{k_m}
= \exp\left( \sum_{n=1}^\infty \frac{z^n}{n}\beta_m(n)\right),
\\
\!\!\!\!\!\!\!\!\!\!\!\!\!
{\mathcal G }(z;X) & = &
\sum_{{{\scriptstyle k_1\geq 0, \ldots, k_m\geq 0,}
\atop\scriptstyle k_1+\ldots+k_m> 0}}{\mathcal C}_+^{(z; m)}({\overrightarrow{k}})
x_1^{k_1}x_2^{k_2}\cdots x_m^{k_m}
= \exp\left( \sum_{n=1}^\infty \frac{(-z)^n}{n}\beta_m(n)\right).
\end{eqnarray}

It is known that the Bell polynomials are very useful in many problems in combinatorics.
We would like to note their
application in multipartite partition problem \cite{Andrews1}.
The Bell polynomials technique can be used for the
calculation ${\mathcal C}_-^{(m)}({\overrightarrow{k}})$ and
${\mathcal C}_+^{(m)}({\overrightarrow{k}})$.
Let
\begin{eqnarray}
\!\!\!\!\!\!\!\!\!\!
\!\!\!\!\!\!\!\!\!\!
&& {\mathcal F}(z;X) := 1 + \sum_{j=1}^\infty {\mathcal P}_j(x_1,x_2, \ldots, x_m)z^j,
\,\,\,\,\,\,\,\,\,\,
{\mathcal P}_j  =  1+ \sum_{{{\scriptstyle k_1\geq 0, \ldots, k_m\geq 0,}
\atop\scriptstyle k_1+\ldots+k_m> 0}}P({\overrightarrow{k}}; j)
x_1^{k_1}\cdots x_m^{k_m},
\label{Fu}
\\
\!\!\!\!\!\!\!\!\!\!
\!\!\!\!\!\!\!\!\!\!
&& {\mathcal G}(z;X)  :=  1 + \sum_{j=1}^\infty {\mathcal Q}_j(x_1,x_2, \ldots, x_m)z^j,
\,\,\,\,\,\,\,\,\,\,
{\mathcal Q}_j  =  1+ \sum_{{{\scriptstyle k_1\geq 0, \ldots, k_m\geq 0,}
\atop\scriptstyle k_1+\ldots+k_m> 0}}Q({\overrightarrow{k}}; j)
x_1^{k_1}\cdots x_m^{k_m}.
\label{Gu}
\end{eqnarray}
Useful expressions for the recurrence relation of the Bell polynomial
$Y_{n}(g_1, g_2, \ldots , g_{n})$
and generating function ${\mathcal B}(z)$ have the forms \cite{Andrews1}:
\begin{eqnarray}
&& Y_{n+1}(g_1, g_2, \ldots , g_{n+1}) = \sum_{k=0}^n  \begin{pmatrix} n\cr k\end{pmatrix}
Y_{n-k}(g_1, g_2, \ldots , g_{n-k})g_{k+1},
\label{B2}
\end{eqnarray}
${\mathcal B}(z) = \sum_{n=0}^\infty Y_n z^n/n! \Longrightarrow
{\rm log}\,{\mathcal B}(z)= \sum_{n=1}^\infty g_n z^n/n!.
$
To verify the last formula we need to differentiate with respect to $z$ and observe
that a comparison of the coefficients of ${z}^n$ in the resulting equation produces an
identity equivalent to (\ref{B2}). From Eq. (\ref{B2}) one can obtain the following explicit
formula for the Bell polynomials (it is known as Faa di Bruno's formula)
\begin{equation}
Y_{n}(g_1, g_2, \ldots , g_{n}) = \sum_{{\bf k}\,\vdash\, n}\frac{n!}{k_1!\cdots k_n!}
\prod_{j=1}^n\left(\frac{g_j}{j!}\right)^{k_j}\!\!.
\end{equation}

Setting $X= (x_1, x_2, \ldots,x_m, 0, 0, \ldots)$; for finite additive manner, then 
the following result holds 
(see for detail \cite{Andrews}):
\begin{eqnarray}
{\mathcal P}_j  & = & \frac{1}{j!}Y_j \left( 0!\beta_m(1),\,\, 1!\beta_m(2)\,\,,
\ldots , \,\,(j-1)!\beta_m(j)\right),
\label{P1}
\\
{\mathcal Q}_j  & = & \frac{1}{(-1)^jj!}Y_j \left( -0!\beta_m(1),\,\,
-1!\beta_m(2)\,\,, \ldots , \,\,-(j-1)!\beta_m(j)\right).
\label{Q1}
\end{eqnarray}
and
\begin{eqnarray}
{\mathcal F}(z;X) & = & 1 + \sum_{j=1}^\infty {\mathcal P}_j(x_1,x_2, \ldots, x_m)z^j
\nonumber \\
& = & 1+ \sum_{j=1}^\infty \frac{z^j}{j!}Y_j \left( 0!\beta_m(1),\,\, 1!\beta_m(2)\,\,,
\ldots , \,\,(j-1)!\beta_m(j)\right),
\\
{\mathcal G}(z;X) & = & 1 + \sum_{j=1}^\infty {\mathcal Q}_j(x_1,x_2, \ldots, x_m)z^j
\nonumber \\
& = & 1 + \sum_{j=1}^\infty \frac{(-1)^jz^j}{j!}Y_j \left( -0!\beta_m(1),\,\,
-1!\beta_m(2)\,\,, \ldots , \,\,-(j-1)!\beta_m(j)\right). 
\end{eqnarray}

\subsection{Restricted specializations}

For some specializations, when $X = (x_1, x_2, \ldots,x_m, 0, 0, \ldots) =
(\underbrace{q,q, ...,q}_m,0,0,\ldots)$ we get
\begin{eqnarray}
{\mathcal F}(z;X) & = &
\prod_{{{\scriptstyle k_1\geq 0, \ldots, k_m\geq 0,}
\atop\scriptstyle k_1+\ldots+k_m> 0}} \left( 1- z q^{k_1+k_2+\cdots + k_m}\right)^{-1}
= \exp\left( - \sum_{n=1}^\infty \frac{z^n}{n}(1-q^{n})^{-m}\right),
\label{F}
\\
{\mathcal G}(z;X) & = &
\prod_{{{\scriptstyle k_1\geq 0, \ldots, k_m\geq 0,}
\atop\scriptstyle k_1+\ldots+k_m> 0}} \left( 1+ z q^{k_1+k_2+\cdots + k_m}\right)
= \exp\left( - \sum_{n =1}^\infty \frac{(-z)^n}{n}(1-q^{n})^{-m}\right).
\label{G}
\end{eqnarray}

{\bf Spectral functions of hyperbolic three-geometry.}
Interesting combinatorial identities may be obtained by applying Euler-Poincar\'e formula to 
graded algebras, for example, to the subalgebras of Kac-Moody algebras 
(see, for example, \cite{Fuks}).
From the point of view of the applications, homologies associated with algebras
$\mg = {\ms}{\ml}(N;{\mathbb C})$ important since they constitute the thechnical basis of the 
proof of the combinatorial identities of Euler-Gauss-Jacobi-MacDonald.

Let us begin by explaining the general lore for the $\mg$-structure on compact groups. 
We recall some results on the Ruelle (Patterson-Selberg type) spectral functions. For details
we refer the reader to \cite{BB,BCST} where spectral
functions of hyperbolic three-geometry were considered in connection with
three-dimensional Euclidean black holes, pure supergravity, and string amplitudes.

Let ${\Gamma}^\gamma \in G=SL(2, {\mathbb C})$ be the discrete group defined by
\begin{eqnarray}
{\Gamma}^\gamma & = & \{{\rm diag}(e^{2n\pi ({\rm Im}\,\vartheta + i{\rm
Re}\,\vartheta)},\,\,  e^{-2n\pi ({\rm Im}\,\vartheta + i{\rm Re}\,\vartheta)}):
n\in {\mathbb Z}\} = \{{\gamma}^n:\, n\in {\mathbb Z}\}\,,
\nonumber \\
{\gamma} & = & {\rm diag}(e^{2\pi ({\rm Im}\,\vartheta + i{\rm
Re}\,\vartheta)},\,\,  e^{-2\pi ({\rm Im}\,\vartheta + i{\rm Re}\,\vartheta)})\,.
\label{group}
\end{eqnarray}
One can construct a zeta function of Selberg-type for the group
${\Gamma}^\gamma \equiv {\Gamma}_{(\alpha, \beta)}^\gamma$ generated by a
single hyperbolic
element of the form ${\gamma_{(\alpha, \beta)}} = {\rm diag}(e^z, e^{-z})$,
where $z= \alpha +i\beta$ for $\alpha, \beta >0$. Actually $\alpha = 2\pi {\rm Im}\,\vartheta$
and $\beta = 2\pi {\rm Re}\,\vartheta$.
The Patterson-Selberg spectral function $Z_{{\Gamma}^\gamma} (s)$ and its logarithm
for ${\rm Re}\, s> 0$ can be attached
to $H^3/{\Gamma}^\gamma$ as follows:
\begin{eqnarray}
Z_{{\Gamma}^\gamma}(s) & := & \prod_{k_1,k_2\geq
0}[1-(e^{i\beta})^{k_1}(e^{-i\beta})^{k_2}e^{-(k_1+k_2+s)\alpha}]\,,
\label{zeta00}
\\
{\rm log}\, Z_{{\Gamma}^\gamma} (s) \!\! & = &
-\frac{1}{4}\sum_{n = 1}^{\infty}\frac{e^{-n\alpha(s-1)}}
{n[\sinh^2\left(\frac{\alpha n}{2}\right)
+\sin^2\left(\frac{\beta n}{2}\right)]}\,.
\label{logZ}
\end{eqnarray}
The zeros of $Z_{\Gamma^\gamma} (s)$ are precisely the set of complex numbers
$
\zeta_{n,k_{1},k_{2}} = -\left(k_{1}+k_{2}\right)+ i\left(k_{1}-
k_{2}\right) \beta/\alpha + 2\pi i n/\alpha,$ with $n \in {\mathbb Z}$.
The magnitude of the zeta-function is bounded
for both ${\rm Re}\,s\geq 0$ and ${\rm Re}\,s\leq 0$, and its growth can be estimated as
\begin{equation}
\big|Z_{\Gamma^\gamma}(s) \big| \leq \Big(\,
\prod_{k_1+k_2\leq
|s|}\,  \e^{|s|\, \ell}\, \Big)\,
\Big(\, \prod_{k_1+k_2\geq
|s|}\, \big(1- \e^{(|s|-k_1-k_2)\, \ell} \big)\,\Big)
\leq C_1\,\e^{C_2\, |s|^3}
\label{estimate}
\end{equation}
for suitable constants $\ell,C_1, C_2$. The first product on the right-hand side of
(\ref{estimate}) gives the exponential growth, while the second product is bounded.
The spectral function $Z_{\Gamma^\gamma} (s)$ is an entire function of
order three and of finite type which can be written as a Hadamard product~\cite{BCST}
\begin{equation}
Z_{\Gamma^\gamma}(s) =
\e^{Q(s)} \
\prod_{\zeta \in {\Sigma}}\,
\Big(\, 1-\frac{s}{\zeta}\, \Big)\,
\exp \Big(\,
\frac{s}{\zeta} + \frac{s^2}{2\zeta^2} +
\frac{s^3}{3\zeta^3}\, \Big)\ ,
\label{Hadamard}
\end{equation}
where $\Sigma$ is the set of zeroes $\zeta := \zeta_{n,k_{1},k_{2}}$
and $Q(s)$ is a polynomial of degree at most three. (The product formula for entire
function (\ref{Hadamard}) is also known as Weierstrass formula  (1876).)

Let us introduce next the Ruelle spectral function ${\mathcal R}(s)$ associated with
hyperbolic three-geometry \cite{BB,BCST}. The function ${\mathcal R}(s)$ is an
alternating product of more complicate factors, each of which is so-called Patterson-Selberg
zeta-functions $Z_{\Gamma^\gamma}$ \cite{Patterson}.
Functions ${\mathcal R}(s)$ can be continued meromorphically to
the entire complex plane $\mathbb C$, poles of ${\mathcal R}(s)$ correspond to zeros of
$Z_{\Gamma^\gamma}(s)$.
\begin{eqnarray}
\prod_{n=\ell}^{\infty}(1- q^{an+\varepsilon})
& = & \prod_{p=0, 1}Z_{\Gamma^\gamma}(\underbrace{(a\ell+\varepsilon)(1-i\varrho(\vartheta))
+ 1 -a}_s + a(1 + i\varrho(\vartheta)p)^{(-1)^p}
\nonumber \\
& = &
\cR(s = (a\ell + \varepsilon)(1-i\varrho(\vartheta)) + 1-a),
\label{R1}
\\
\prod_{n=\ell}^{\infty}(1+ q^{an+\varepsilon})
& = &
\prod_{p=0, 1}Z_{\Gamma^\gamma}(\underbrace{(a\ell+\varepsilon)(1-i\varrho(\vartheta)) + 1-a +
i\sigma(\vartheta)}_s
+ a(1+ i\varrho(\vartheta)p)^{(-1)^p}
\nonumber \\
& = &
\cR(s = (a\ell + \varepsilon)(1-i\varrho(\vartheta)) + 1-a + i\sigma(\vartheta))\,,
\label{R2}
\end{eqnarray}
\begin{eqnarray}
\prod_{n=\ell}^{\infty}(1-q^{an+ \varepsilon})^{bn} & = &
\cR(s=(a\ell + \varepsilon)(1-i\varrho(\vartheta))+1-a)^{b\ell}
\nonumber \\
& \times &
\!\!\!
\prod_{n=\ell+1}^{\infty}
\cR(s=(an + \varepsilon)(1-i\varrho(\vartheta))+1-a)^{b}\,,
\label{RU1}
\\
\prod_{n=\ell}^{\infty}(1+q^{an+ \varepsilon})^{bn} & = &
\cR(s=(a\ell + \varepsilon)(1-i\varrho(\vartheta))+1-a+ i\sigma(\vartheta))^{b\ell}
\nonumber \\
& \times &
\!\!\!
\prod_{n=\ell+1}^{\infty}
\cR(s=(an + \varepsilon)(1-i\varrho(\vartheta))+1-a+ i\sigma(\vartheta))^{b}\,,
\label{RU2}
\end{eqnarray}
being $q\equiv e^{2\pi i\vartheta}$, $\varrho(\vartheta) =
{\rm Re}\,\vartheta/{\rm Im}\,\vartheta$,
$\sigma(\vartheta) = (2\,{\rm Im}\,\vartheta)^{-1}$,
$a$ is a real number, $\varepsilon, b\in {\mathbb C}$, $\ell \in {\mathbb Z}_+$.

Obviously, 
\begin{eqnarray}
\beta_m(n) & = & \prod_{j=1}^m(1- q^{jn})^{-1}
\equiv \prod_{j = 1}^\infty(1-q^{jn})^{-1}\prod_{j = m+1}^\infty(1-q^{jn}) 
\nonumber \\
& = & 
\frac{\cR(s= n(m+1)(1-i\varrho(\vartheta))+1-n)}
{\cR(s= n(1-i\varrho(\vartheta))+1-n)}
\label{beta}
\end{eqnarray}
and
\begin{eqnarray}
{\mathcal F}(z;X) & = &
\prod_{{{\scriptstyle k_1\geq 0, \ldots, k_m\geq 0,}
\atop\scriptstyle k_1+\ldots+k_m> 0}} \left( 1- z q^{k_1+k_2+\cdots + k_m}\right)^{-1}
\nonumber \\
& \stackrel{by \,\, (\ref{F})}{=\!=\!=\!=} &
\times\exp\left( - \sum_{n=1}^\infty 
\frac{z^n\cR(s= -in\varrho(\vartheta)(m+1)+nm+1)}
{n\,\cR(s = -in\varrho(\vartheta)+ 1)}\right),
\label{F1}
\\
{\mathcal G}(z;X) & = &
\prod_{{{\scriptstyle k_1\geq 0, \ldots, k_m\geq 0,}
\atop\scriptstyle k_1+\ldots+k_m> 0}} \left( 1+ z q^{k_1+k_2+\cdots + k_m}\right)
\nonumber \\
& \stackrel{by \,\, (\ref{G})}{=\!=\!=\!=} &
\times\exp\left( - \sum_{n =1}^\infty 
\frac{(-z)^n\cR(s= -in\varrho(\vartheta)(m+1)+nm+1)}
{n\,\cR(s = -in\varrho(\vartheta)+ 1)}\right)\,.
\label{G1}
\end{eqnarray}
Also series for $\cF(z; X)$ and $\cG(z; X)$ have the forms (\ref{F}) and (\ref{G}) 
correspondingly with $\beta_m(n)$ is given by Eq. (\ref{beta}).
\\

{\bf Example:} Let us calculate $\cP_2$ coefficient. With the help of recurrence relation 
(\ref{B2}) we obtain
\begin{eqnarray}
\!\!\!\!\!\!\!\!
2\cP_2 & = & (Y_2(\beta_m(1), \beta_m(2)) = Y_2(\beta_m(1)^2 + \beta_m(2))
= 
\prod_{j=1}^m(1-q^{j2})^{-1} + \prod_{j=1}^m(1-q^{j2})
\nonumber \\
& = &
\frac{\cR(s= 2(m+1)(1-i\varrho(\vartheta))-1)^2 + \cR(s= 2(1-i\varrho(\vartheta)-1))^2}
{\cR(s= 2(m+1)(1-i\varrho(\vartheta))-1)\cdot\cR(s= 2(1-i\varrho(\vartheta)-1))} .
\end{eqnarray}

\begin{remark}
In the simple case when $X = (q, 0, 0, \ldots )$ and $z=1$ we have
$
{\mathcal F}(1;q)^{a_k} = \prod_{k \geq 0}(1-q^k)^{-a_k}. 
$
There are some expansions which are differ from power series expansions that are 
useful in imperical studies. Indeed the following result holds (see also \cite{Andrews})
\begin{equation} 
\prod_{k=1}^\infty (1-q^k)^{-a_k}  =  1+ \sum_{k=1}^\infty \cB_k q^k,
\label{prod1}
\end{equation}
$k\cB_k  =  \sum_{j=1}^k\cD_j\cB_{k-j} q^k,\, \cD_j = \sum_{d\vert j}da_d$.
\label{prod1}
Here $a_k$ and $\cB_k$ are integers. Note that if either sequance $a_k$ or $\cB_k$ is 
given, the other is uniquely determined by $cB_k$ and $\cD_j$. 
\end{remark}

\subsection{The infinite hierarchy}

Setting $z q^{k_{1}+\ldots + k_{m}}q^n = z\varOmega_{{\overrightarrow{k}}}q^{n_1}$
with $\varOmega_{{\overrightarrow{k}}}=
q^{k_{1}+\ldots +k_{m}}$\, (${\overrightarrow{k}} =
\left( k_{1},\ldots ,k_{m}\right))$ we get
\begin{equation}
G_{1}\left( z\varOmega_{{\overrightarrow{k}}};q\right) :=\prod_{n_1=0}^{\infty}
(1-z\varOmega_{{\overrightarrow{k}}}q^{n_1}) =
(1-z\varOmega_{{\overrightarrow{k}}})\cdot
{\mathcal R}(s= (1 + {\overline \Omega}(z\varOmega_{{\overrightarrow k}}))
(1- i\varrho{\vartheta}))\,,
\label{G1}
\end{equation}
where ${\overline \Omega}(z\varOmega_{{\overrightarrow k}})\equiv 
{\rm log} (z \varOmega_{\overrightarrow k})/2i\pi\vartheta$.
Therefore the infinite products can be factorized as 
\begin{equation}
\prod_{k_m=0}^\infty\prod_{k_{m-1}=0}^\infty\cdots\prod_{k_1=0}^\infty\prod_{n_1=0}^\infty
(1-z\varOmega_{{\overrightarrow{k}}}q^n_1)=
\prod_{{\overrightarrow{k}}\geq {\overrightarrow{0}}} G_{1}\left(
z\varOmega_{{\overrightarrow{k}}};q\right).
\end{equation}
We can treat this factorization as a product of $m$ copies, each of them is
$G_{1}\left( z\varOmega_{{\overrightarrow{k}}};q\right)$ and corresponds
to a free two-dimensional conformal field theory. The next step of the iterative 
loop becomes the Jackson (convergent) double infinite product $G_2(z; q, p)$
\cite{Jackson}
\begin{eqnarray}
G_2( z\varOmega_{{\overrightarrow{k}}};q, p) & = & 
\prod_{n_2,n_1= 0}^\infty(1- z\varOmega_{{\overrightarrow{k}}}q^{n_1 + n_2})
=
\prod_{n_2 = 0}^\infty (1-z\varOmega_{{\overrightarrow{k}}}q^{n_2})
\nonumber \\
& \times & 
{\mathcal R}(s= (1 + {\overline \Omega}(z\varOmega_{{\overrightarrow k}}q^{n_2})
(1- i\varrho{\vartheta}))
\label{G2}
\end{eqnarray}

For the product (\ref{G2}) two first order $q$- and $p$-equations 
take the forms \cite{Spiridonov}
\begin{eqnarray}
\frac{G_2( \varOmega_{{\overrightarrow{k}}};q, p)}
{G_2( q\varOmega_{{\overrightarrow{k}}};q, p)} & = & 
G_1( \varOmega_{{\overrightarrow{k}}};p),
\,\,\,\,\,\,\,\,\,\,\,\,\,\,\,\,\,\,\,\,\,\,\,\,\,
\frac{G_2( \varOmega_{{\overrightarrow{k}}};q, p)}
{G_2(p \varOmega_{{\overrightarrow{k}}};q, p)} = 
G_1( \varOmega_{{\overrightarrow{k}}};q)\,,
\label{G21}
\\
\!\!\!\!\!\!\!\!\!\!\!\!\!\! 
\frac{G_2(qp(q \varOmega_{{\overrightarrow{k}}})^{-1};q, p)}
{G_2(qp (\varOmega_{{\overrightarrow{k}}})^{-1};q, p)} & = & 
G_2(p (\varOmega_{{\overrightarrow{k}}})^{-1};p)\,,
\,\,\,\,\,\,\,\,\,\,
\frac{G_2(qp (p\varOmega_{{\overrightarrow{k}}})^{-1};q, p)}
{G_2(qp( \varOmega_{{\overrightarrow{k}}})^{-1};q, p)} = 
G_2(q( \varOmega_{{\overrightarrow{k}}})^{-1};q).
\label{G22} 
\end{eqnarray}

Symmetry roperties of Jackson double infinite product $G_2(z;q,p)$ analogous to 
(modular) properties of the standard elliptic gamma functions.
For $z\in {\mathbb C}^\ast$ the order one $\Gamma_1$ and double (i.e., the order two) $\Gamma_2$ 
standard elliptic gamma functions have the forms
\begin{eqnarray}
\Gamma_1 (z;q,p) & = & \prod_{n_1,n_2 = 0}^\infty\left(\frac{1-z^{-1}q^{n_1+1}p^{n_2+1}}
{1- zq^{n_1}p^{n_2}}\right),
\nonumber \\
\Gamma_2(z; q,p,t) & = & 
\!\! \prod_{n_1,n_2,n_3 = 0}^\infty
(1- z^{-1}q^{n_1+1}p^{n_2+1}t^{n_3+1})(1- zq^{n_1}p^{n_2}t^{n_3}).
\end{eqnarray}
The double elliptic gamma function $\Gamma_2$ has the following interesting modular 
properties:
\begin{eqnarray}
\Gamma_2(z; a, b, c) & = & \Gamma_2(z/a; -1/a,b/a,c/a)\cdot\Gamma_2(z/b; a/b,-1/b,c/b)
\cdot\Gamma_2(z/c; a/c,b/c,-1/c)
\nonumber \\
&\times &
\!\!{\rm exp}\left(\frac{i\pi}{12}B_{44}(z; a,b,c)\right)\,, 
\end{eqnarray}
where $B_{44}$ is given by
\begin{equation}
B_{44}(z; a,b,c) = \lim_{\stackrel{x\rightarrow 0}{}}\frac{d^4}{dx^4}\left(\frac{x^4e^{zx}}
{(e^{ax}-1)(e^{bx}-1)(e^{cx}-1)}\right) 
\end{equation}
and $2i\pi a = {\rm log}\,q,\,2i\pi b = {\rm log}\,p,\,2i\pi c = {\rm log}\,t$.
In the case when $q=p=t$ we get
\begin{eqnarray}
\Gamma_1 (z;q,q) & = & \prod_{n_2 = 0}^\infty\prod_{n_1 = 0}^\infty
\left(\frac{1-z^{-1}q^{n_1+1}p^{n_2+1}}
{1- zq^{n_1}p^{n_2}}\right) = \prod_{n_2 = 0}^\infty
\left(\frac{1-z^{-1}q^{n_2+2}}{1-zq^{n_2}}\right)
\nonumber \\
& \times &
\left(\frac{\cR(s= (n_2+ {\overline \Omega}(z^{-1}; \vartheta) +2)(1-i\varrho(\vartheta))}
{\cR(s = (n_2+ {\overline \Omega}(z; \vartheta))(1-i\varrho(\vartheta))}\right),
\end{eqnarray}
where ${\overline \Omega}(z^{\pm 1}; \vartheta) = \pm {\rm log}\,z/2\pi i \vartheta$.
\begin{eqnarray}
\Gamma_2(z; q,q,q) & = & \!\!\prod_{n_2,n_3 = 0}^\infty\prod_{n_1=0}^\infty
(1- z^{-1}q^{n_1+n_2+n_3 +3})(1- zq^{n_1+n_2 +n_3})
\nonumber \\
& = & \!\! \prod_{n_2,n_3= 0}^\infty (1-z^{-1}q^{n_2+n_3+ 2})(1-zq^{n_2+n_3})
\nonumber \\
&\times & 
\!\!\cR(s = (n_2+n_3+{\overline \Omega}(z^{-1}; \vartheta)+3)(1-i\varrho(\vartheta)))
\nonumber \\
& \times &
\!\!\cR(s = (n_2+n_3+{\overline \Omega}(z; \vartheta)+1)(1-i\varrho(\vartheta))).
\end{eqnarray}

\section{The quantum group invariants}
\label{Quantum}

In this Sect. we view correlators in a CS theory as generating series of quantum group invariants 
weighted by S-functions. The quantum group invariants can be defined over any semi-simple Lie 
algebra $\mathfrak g$. In the $SU(N)$ Chern-Simons gauge theory we study the quantum 
${\mathfrak s}{\mathfrak l}_N$ invariants, which can be identified as the so-called colored 
HOMFLY polynomials.

One important corollary of the LMOV conjecture is the possibility to express 
a Chern-Simons partition function as an infinite product. In this article we derive such a product. 
During the calculations we use the characters of the symplectic groups. The latter were
found by Weyl \cite{Weyl} using a transcendental method
(based on integration over the group manifold). However the appropriate characters may 
also be obtained by algebraic methods \cite{Littlewood44}. Following \cite{Fauser10} we have 
used algebraic methods. This allows to exploit the Hopf algebra methods to determine
(sub)group branching rules and the decomposition of tensor products. 
 
The motivation for studying an infinite-product formula, associated to topological string 
partition functions, based on a guess on the modular property of partition function, 
stimulated by properties of S-functions.

{\bf Review of basic tools.}
To derive the infinite-product formula, we need some preliminary material. 
First of all we denote by $\cY$ the set of all Young diagrams. Let $\chi_A$ 
be the character of the irreducible
representation of the symmetric group labeled by a partition $A$. Given a partition $\mu$,
define $m_j = \textrm{card} (\mu_k=j; k\geq 1)$. (The order of the conjugate class of type
$\mu$ is given by: $\mathfrak{z}_\mu = \prod_{j\geq1} j^{m_j} m_j!.$)
The symmetric power functions of a given set of variables $X=\{x_j\}_{j\geq 1}$
are defined as the direct limit of the Newton polynomials:
$p_n(X) = \sum_{j\geq1} x_j^n, \, \, p_\mu(X) = \prod_{i\geq 1} p_{\mu_i}(X),$
and we have the following formulae which determine the Schur function and the orthogonality
property of the character
\begin{equation}
s_A(X) = \sum_{\mu} \frac{ \chi_A(C_\mu) }{\mathfrak{z}_\mu} p_\mu(X), \,\,\,\,\,\,\,\,\,
\sum_{\mu} \frac{ \chi_A(C_\mu) \chi_B(C_\mu) }{ \mathfrak{z}_\mu } = \delta_{A,B}\,.
\end{equation}
where $C_\mu$ denotes the conjugate class of the symmetric 
{group $S_{\vert \mu \vert}$} 
corresponding to partition $\mu$ (for details see Sect. 3 of [33]).

Given $X= \{x_i\}_{i\geq 1}$, $Y=\{y_j\}_{j\geq 1}$, define
$
X\ast Y = \{x_i\cdot y_j\}_{i\geq 1, j\geq 1}.
$
We also define $X^d = \{ x_i^d\}_{i\geq 1}$.
The $d$-th Adams operation of a Schur function is given by $s_A(X^d)$. 
(An Adams operation is type of algebraic construction; the basic idea of this operation 
is to implement some fundamental identities in S-functions. In particular, $s_A(X^d)$ 
means operation of a power sum on a polynomial.)
We use the following conventions for the notation:
\begin{itemize}
\item{} 
$\cL$ will denote a link and $L$ the number of components in $\cL$.
\item{}
The irreducible $U_q(\mathfrak{sl}_N)$ module associated to $\mathcal L$ will be labeled by
their highest   weights, thus by Young diagrams. We usually denote it by a vector form
$\overrightarrow{A}=(A^1,\ldots,A^L)$.
\item{}
Let $\overrightarrow{X} =(x_1,\ldots,x_L)$ be a set of $L$  variables,
each of which is associated to  a component of $\mathcal L$ and
$\overrightarrow{\mu} = (\mu^1,\ldots,\mu^L)\in\cY^L$ be a tuple of $L$ partitions. We write:
$$
[\overrightarrow{\mu}] = \prod_{\alpha=1}^L [\mu^\alpha],
\,\,\,\,\,\,\, \mathfrak{z}_{\overrightarrow{\mu}} = \prod_{\alpha=1}^L \mathfrak{z}_{\mu^\alpha},
\,\,\,\,\,\,\, \chi_{\overrightarrow{A}}(C_{\overrightarrow{\mu}}) =
\prod_{\alpha=1}^L \chi_{A^\alpha}(C_{\mu^\alpha}),
$$
$$
s_{\overrightarrow{A}}(\overrightarrow{X}) =
\prod_{\alpha=1}^L s_{A^\alpha}(x_\alpha), \,\,\,\,\,\,\,
p_{\mu }(X)=\overset{\ell (\mu )}{\prod_{i=1}}p_{\mu _{i}}(X),\,\,\,\,\,\,\,
p_{\overrightarrow{\mu}}(\overrightarrow{X}) = \prod_{\alpha=1}^L p_{\mu^\alpha}(x_\alpha).
\nonumber
$$
\end{itemize}

{\bf The case of links and a knot.}
The quantum $\mathfrak{sl}_N$ invariant for the irreducible module $V_{A^1},\ldots,V_{A^L}$,
labeled by the corresponding partitions $A^1,\ldots, A^L$,  can be
identified as the HOMFLY invariants for the link decorated by $Q_{A^1},\ldots,Q_{A^L}$.
The quantum $\mathfrak{sl}_N$ invariants of the link is given by
$
P_{\overrightarrow{A}}(\mathcal{L}; q,t) =
\mathcal{H} (\mathcal{L}\star \otimes_{\alpha=1}^L Q_{A^\alpha} ).
$
The colored HOMFLY polynomial of the link $\mathcal L$ can be defined by \cite{Zhu}
\begin{equation}
P_{\overrightarrow{A}} = q^{-\sum_{\alpha = 1}^Lk_{A^\alpha}\omega({\mathcal K}_\alpha)}
t^{-\sum_{\alpha = 1}^L \vert A^\alpha\vert \omega({\mathcal K}_\alpha)}
\langle \mathcal{L}\star \otimes_{\alpha=1}^L Q_{A^\alpha} \rangle\,,
\end{equation}
where $\omega({\mathcal K}_\alpha)$ is the number of the $\alpha$-component
${\mathcal K}_\alpha$ of $\mathcal L$
and the bracket $\langle \mathcal{L}\star \otimes_{\alpha=1}^L Q_{A^\alpha} \rangle$ denotes
the framed HOMFLY polynomial of the satellite link
$\mathcal{L}\star \otimes_{\alpha=1}^L Q_{A^\alpha}$.
We can define the following invariants:
\begin{equation}\label{definition of W_mu}
\textsf{W}_{\overrightarrow{\mu}}(\mathcal{L};q,t) =
\sum_{\overrightarrow{A} =(A^1,\ldots,A^L) } \bigg( \prod_{\alpha=1}^L
\chi_{A^\alpha}(C_{\mu^\alpha} ) \bigg)
P_{\overrightarrow{A}}(\mathcal{L};q,t)\,.
\end{equation}

The Chern-Simons partition function $\textsf{W}^{SL}_{CS}({\mathcal L};q,t)$ and the free energy
$F({\mathcal L};q,t)$ of the link ${\mathcal L}$ are the following generating series of
quantum group invariants weighted by Schur functions $s_{\overrightarrow{A}}$ and by the
invariants $\textsf{W}_{\overrightarrow{\mu}}$:
\begin{eqnarray}
\textsf{W}^{SL}_{CS}({\mathcal L};q,t) & = &
1+ \sum_{\overrightarrow{A}} P_{\overrightarrow{A}} ({\mathcal L}; q,t)
s_{\overrightarrow{A}}(\overrightarrow{X}) =
1+ \sum_{\overrightarrow{\mu} }
\frac{ \textsf{W}_{\overrightarrow{\mu}}
({\mathcal L};q,t) }{{\mathfrak z}_{\overrightarrow{\mu}} }
p_{\overrightarrow{\mu}}(\overrightarrow{X}) \,,
\label{CS-A}
\\
F({\mathcal L};q,t) & = & \log \textsf{W}_{CS}({\mathcal L};q,t)
= \sum_{\overrightarrow{\mu}}
\frac{ F_{\overrightarrow{\mu}}({\mathcal L};q,t) }{{\mathfrak z}_{\overrightarrow{\mu}}}
p_{\overrightarrow{\mu}}(\overrightarrow{X}) \,.
\end{eqnarray}

{\bf From summations to infinite products.}
The Chern-Simons theory has been conjectured to be equivalent to a topological string theory
$1/N$ expansion in physics. This duality conjecture builds a fundamental connection in
mathematics. On the one hand, Chern-Simons theory leads to the
construction of knot invariants; on the other hand, topological string theory gives rise to
Gromov-Witten theory in geometry.

The Chern-Simons/topological string duality conjecture identifies
the generating function of Gromov-Witten invariants as
Chern-Simons knot invariants \cite{OV}.  Based on these thoughts,
the existence of a sequence of integer invariants is conjectured
\cite{OV, LMV} in a similar spirit to  Gopakumar-Vafa
setting \cite{GV}, which provides an essential evidence of the
duality between Chern-Simons theory and topological string theory.
This integrality conjecture is called the LMOV
conjecture. One important corollary of the LMOV conjecture is to
express Chern-Simons partition function as an infinite product
derived in this article. The motivation of studying such an
infinite-product formula is based on a guess on the modularity
property of topological string partition function.

Based on LMOV conjecture the infinite product formulae for the case of links, 
$\textsf{W}_{CS}^{SL}({\mathcal L};q,t; \overrightarrow{X})$ and a knot 
$\textsf{W}_{CS}^{SL}(\cK;q,t;X)$ are given by \cite{Liu2,LiuPeng}  
\begin{eqnarray}
\textsf{W}_{CS}^{SL}(\cK;q,t;X) & = &   
\prod_{\mu}\prod_{Q\in {\mathbb Z}/2} \,\, \prod_{m=1}^\infty \;
\prod_{k = -\infty}^\infty\;
\big \langle 1- q^{k+m}t^Q   X^\mu
\big \rangle^{-m\, {n}_{\mu;\,g,Q}}\,
\label{SP2}
\\
\textsf{W}_{CS}^{SL}({\mathcal L};q,t; \overrightarrow{X}) & = &
\prod_{\overrightarrow{\mu}}\,
\prod_{Q\in {\mathbb Z}/2} \, \prod_{m=1}^\infty\prod_{k= -\infty}^\infty\,
\big\langle 1-  q^{k+m}t^Q \, \overrightarrow{X}\big\rangle^{-mn_{\overrightarrow{\mu};\,g,Q}}.
\label{SP1}
\end{eqnarray} 
Here $\overrightarrow{\mu}=(\mu^1,\ldots,\mu^L)$, the length of $\mu^i$ is $\ell_i$, 
$\overrightarrow{X}=(x_1,\ldots, x_L)$, and 
${n}_{\mu;\,g,Q}$ are invariants related to the integer invariants in the LMOV
conjecture. For a given $\mu$, ${n}_{\mu;g,Q}$ vanish for sufficiently 
large $|Q|$ due to the vanishing property of $n_{\mu;\,g,Q}$. The products involving $Q$ 
and $k$ are finite products for a fixed partition $\mu$.

The symmetric product $\big \langle 1- q^{k+m}t^Q   X^\mu \big \rangle$ and the generalized 
symmetric product $\big\langle 1-  q^{k+m}t^Q \, \overrightarrow{X}\big\rangle$  in Eqs. 
(\ref{SP2}) and (\ref{SP1}), respectively, are defined by the formulae \cite{LiuPeng}
\begin{eqnarray}
\big\langle 1- \psi\, X^\mu \big\rangle & = & 
\prod_{ x_{i_1},\ldots, x_{i_{\ell(\mu)} } }
\Big( 1- \psi\, x_{i_1}^{\mu_1} \cdots x_{i_{\ell(\mu)}}^{\mu_{\ell(\mu)}} \Big)
\label{Angle1},
\\
\big\langle 1-\psi\, \overrightarrow{X}\big\rangle & = &   
\prod_{\alpha=1}^L\prod_{i_{\alpha, 1},\ldots, i_{\alpha, \ell_\alpha}}
\Big(1- \psi\prod_{\alpha=1}^L
\big((x_\alpha)_{ i_{\alpha,1} }^{\mu^\alpha_1} \cdots
(x_\alpha)_{ i_{\alpha, \ell_\alpha} }^{\mu^\alpha_{\ell_\alpha}}\big)\Big)\,.
\label{Angle2}
\end{eqnarray}
where $\psi$ is a generic variable. Because of symmetry $q \rightarrow q^{-1}$, we have
\
\begin{eqnarray}
&&
\prod_{k=-\infty}^\infty(1-q^{k+m}t^QX^\mu)^{-mn_{\overrightarrow{\mu};\,g,Q}}  = 
(1-q^{m}t^QX^\mu)^{-mn_{\overrightarrow{\mu};\,g,Q}}\cdot
\prod_{k= 1}^\infty(1-q^{k+m}t^QX^\mu)^{-2mn_{\overrightarrow{\mu};\,g,Q}}
\nonumber \\
&&
=  
(1-q^{m}t^QX^\mu)^{-mn_{\overrightarrow{\mu};\,g,Q}}\cdot
\cR( s = (1+ \overline{\Omega}(q^mt^Q\Omega_{X^\mu}))(1-i\varrho(\vartheta))))
^{-2mn_{\overrightarrow{\mu};\,g,Q}}\,,
\end{eqnarray}

where $\overline{\Omega}(q^mt^Q\Omega_{X^\mu})\equiv {\rm log} (q^mt^Q\Omega_{X^\mu})/
2i\pi\vartheta, X^\mu \equiv x_{i_1}^{\mu_1}\cdots x_{i_{\ell(\mu)}}^{\mu_{\ell(\mu)}}.$ 
Therefore, $\textsf{W}_{CS}^{SL}(\cK;q,t; X^\mu)$ and 
$\textsf{W}_{CS}^{SL}({\mathcal L};q,t; \overrightarrow{X})$ take the form
\begin{eqnarray}
\textsf{W}_{CS}^{SL}(\cK;q,t; {X^\mu}) & = & \prod_{\mu}\prod_{Q\in {\mathbb Z}/2} \,\, 
\prod_{ x_{i_1},\ldots, x_{i_{\ell(\mu)} }} \prod_{m=1}^\infty 
(1-q^{m}t^QX^\mu)^{-mn_{\overrightarrow{\mu};\,g,Q}}
\nonumber \\
& \times &
\cR( s = (1+ \overline{\Omega}(q^mt^Q\Omega_{X^\mu}))(1-i\varrho(\vartheta))))
^{-2mn_{\overrightarrow{\mu};\,g,Q}}
\label{Knot}
\\
\textsf{W}_{CS}^{SL}({\mathcal L};q,t;X) & = &
\prod_{\mu}\prod_{Q\in {\mathbb Z}/2} \,\, \prod_{\alpha=1}^L\prod_{i_{\alpha, 1},
\ldots, i_{\alpha, \ell_\alpha}}\prod_{m=1}^\infty 
(1-q^{m}t^Q\overrightarrow{X})^{-mn_{\overrightarrow{\mu};\,g,Q}}
\nonumber \\
& \times &
\cR( s = (1+ \overline{\Omega}(q^mt^Q\Omega_{\overrightarrow{X}}))(1-i\varrho(\vartheta))))
^{-2mn_{\overrightarrow{\mu};\,g,Q}}\,.
\label{Link}
\end{eqnarray}
For further simplification we may use Eq. (\ref{RU1}).
Similar calculations in the case of Kauffman polynomials, relative to the orthogonal group, 
can be found in a recent paper \cite{BC}.

{\bf Symmetry and modular properties in infinite-product structure.}
\label{Symmetry}
In this section we discuss a basic symmetric property of infinite-product structure
obtained from the LMOV partition function.
For this reason we can use functional equations for the
spectral Ruelle functions (\ref{R1})-- (\ref{RU2}):
\begin{eqnarray}
\!\!\!\!\!\!\!\!\!\!
&&
\!\!\!\!\!\!\!\!\!\!\!\!\!\!
{\mathcal R}(s= ({z}+b)(1-i\varrho(\vartheta))+i\sigma(\vartheta))\cdot
{\mathcal R}(s= -(1+{z}+b)(1-i\varrho(\vartheta))+i\sigma(\vartheta))
\nonumber \\
\!\!\!\!\!\!\!\!\!\!
=
&&
\!\!\!\!\!\!\!
q^{-{z}b-b(b+1)/2}
{\mathcal R}(s= -{z}(1-i\varrho(\vartheta))+i\sigma(\vartheta))\cdot
{\mathcal R}(s= (1+{z})(1-i\varrho(\vartheta))+i\sigma(\vartheta))
\nonumber \\
\!\!\!\!\!\!\!\!\!\!
=
&&
\!\!\!\!\!\!\!
q^{-{z}(b-1)-b(b+1)/2}
{\mathcal R}(s= (1-{z})(1-i\varrho(\vartheta))+i\sigma(\vartheta))\cdot
{\mathcal R}(s= {z}(1-i\varrho(\vartheta))+i\sigma(\vartheta)).
\label{DE3}
\end{eqnarray}

The simple case $b=0$ in Eq. (\ref{DE3}) leads to the symmetry
$\vartheta \rightarrow -\vartheta$, i.e the symmetry $q\rightarrow q^{-1}$.

There is also the following symmetry about $\mu$ and $Q$
$
n_{\mu;\, g,-Q} = (-1)^{\ell(\mu)} n_{\mu;\, g, Q},
$
which can be interpreted as the rank-level duality of the $SU(N)_k$ and $SU(k)_N$
Chern-Simons gauge theories \cite{LiuPeng}. Rank-level duality is essentially a symmetry of
quantum group invariants relating a labeling color to its transpose \cite{LiuPeng}. It can be
expressed using symmetry about $\mu$, $Q$, and modularity properties of Ruelle functions
as follows:
$
W_{ A^t}( s^{-1}, -v ) = W_{A} (s, v)\,,
$
where $s=q^{1/2}$, $v=t^{1/2}$. The stronger version is \cite{LiuPeng,Zhu,CLPZ}:
$
W_{A^t}( s^{-1}, v) = (-1)^{|A|} W_A (s, v),\,
W_A(s, -v) = (-1)^{|A|} W_A(s, v).
$

\end{document}